\documentclass[aps,prb,preprint,groupedaddress]{revtex4-1}

\usepackage{graphicx}
\usepackage{amsmath}
\usepackage{natbib}

\usepackage{color}

\usepackage{ulem}

\begin{document}

\title{The Momentum Distribution of Liquid $^4$He}

\author{T.R. Prisk}
\email[]{timothy.prisk@nist.gov}
\affiliation{Center for Neutron Research, National Institute of Standards and Technology, Gaithersburg, MD 20899-6100, USA}

\author{M.S. Bryan}
\author{P.E. Sokol}
\affiliation{Department of Physics, Indiana University, Bloomington, IN 47408, USA}

\author{G.E. Granroth}
\affiliation{Neutron Data Analysis and Visualization Division, Oak Ridge National Laboratory, Oak Ridge, TN 37831, USA}

\author{S. Moroni}
\affiliation{CNR-IOM DEMOCRITOS, Istituto Officina dei Materiali, and SISSA Scuola Internazionale Superiore di Studi Avanzati, Via Bonomea 265, I-34136 Trieste, Italy}

\author{M. Boninsegni}
\affiliation{Department of Physics, University of Alberta, Edmonton, Alberta T6G 2E1, Canada}

\date{\today}

\begin{abstract}
We report high-resolution neutron Compton scattering measurements of liquid $^4$He under saturated vapor pressure.  There is excellent agreement between the observed scattering and \emph{ab initio} predictions of its lineshape.  Quantum Monte Carlo calculations predict that the Bose condensate fraction is zero in the normal fluid, builds up rapidly just below the superfluid transition temperature, and reaches a value of approximately $7.5\%$ below 1 K.  We also used model fit functions to obtain from the scattering data empirical estimates for the average atomic kinetic energy and Bose condensate fraction.  These quantities are also in excellent agreement with \emph{ab initio} calculations.  The convergence between the scattering data and Quantum Monte Carlo calculations is strong evidence for a Bose broken symmetry in superfluid $^4$He.
\end{abstract}

%\pacs{}

\maketitle

\newpage

\section{Introduction}\label{Introduction}

Bose condensation plays an important role in contemporary quantum many-body physics.  It is generally believed to provide the microscopic basis for superfluidity in bulk liquid $^4$He\cite{PSbook, PN2, Anderson}.  On this view, the local order parameter of the normal-to-superfluid phase transition is a macroscopic wavefunction describing the motion of the condensate.  The spontaneously broken gauge symmetry produces many fundamental properties of superfluid $^4$He, such as its two-fluid hydrodynamics, critical exponents, and quantization of circulation.  Recent interest in Bose condensation encompasses a broad range of topics\cite{BEC}, including dilute atomic gases\cite{GNZ}, solid state excitations\cite{Snoke, Balili, Kasprzak, Nikuni, Demokritov, Deng}, non-linear optical systems\cite{Klaers, Carusotto}, neutron stars\cite{Graber}, and gravitation\cite{vanZoest, ODell, Peres}.  Bulk superfluid $^4$He represents the strongly interacting limit of Bose-condensed systems due to the steeply repulsive core of its interatomic potential.

Experimental tests of Bose broken symmetry in superfluid $^4$He are therefore a subject of fundamental interest in quantum many-body physics.  The only known physical property of liquid $^4$He that can provide direct information about the existence and magnitude of its Bose condensate is the atomic momentum distribution $n(k)$.  Hohenberg and Platzman proposed that neutron Compton scattering be used to detect the Bose condensate in superfluid $^4$He\cite{HohenbergPlatzman}.  Their proposal stimulated many experimental efforts to determine the Bose condensate fraction $n_0$ throughout the phase diagram of $^4$He\cite{Martel, Svensson, SilverSokol, Glyde, Andreani}.

Several groups have performed measurements of $n(k)$ using the high fluxes of epithermal neutrons available at modern spallation sources.  This includes the MARI\cite{Azuah1, Azuah2, GAS, MariP, Diallo1, Diallo2}, PHOENIX\cite{Sosnick, SosnickLong, Snow, SnowLong, HerwigP}, and eVS\cite{Mayers} spectrometers.  There is both convergence and divergence in the results of these studies.  All groups find that the condensate fraction $n_0$ is zero in the normal fluid and solid phases.  They also obtain physically plausible values for $n_0$ as a function of temperature and pressure.  However, their empirical estimates for the condensate fraction $n_0$ are in quantitative disagreement.  For example, the MARI, PHOENIX, and eVS groups report that the ground state value of $n_0$ under zero applied pressure is $n_0 = (7.25 \pm 0.75)\%, (10 \pm 1.25)\%, (15 \pm 4)\%$, respectively.  The reason for this discrepancy has not been clarified in the literature.

In this paper, we present a new high-resolution neutron Compton scattering study of liquid $^4$He under saturated vapor pressure.  The measurements were carried out using the Wide Angular Range Chopper Spectrometer (ARCS) at the Spallation Neutron Source\cite{mason2006spallation}.  To interpret the scattering data, we performed Quantum Monte Carlo (QMC) calculations of the atomic momentum distribution $n(k)$.  We accounted for corrections to the Impulse Approximation (IA) by using the theories of Silver\cite{Silver1, Silver2} and Carraro-Koonin\cite{CK, Moroni2}.  There is excellent agreement between the ARCS data set and QMC predictions when the latter theory is used.  In particular, the condensate fraction $n_0$ is zero in the normal fluid, becomes finite in the critical region below $T_\lambda$, and reaches a value of $7.5\%$ at 1.09 K.
  
The paper is organized as follows.  In Section II, we review the conceptual framework of neutron Compton scattering and describe the theoretically expected scattering.  Section III provides the details of the experimental setup, instrument characterization, and data treatment.  Section IV presents the experimental data without any reference to theoretical models.  In Section V, we compare our experimental data to theoretical predictions.  We also use model fit functions to obtain empirical estimates for the average atomic kinetic energy $\langle E_K\rangle$ and condensate fraction $n_0$ as a function of temperature.  The scattering data collected by the PHOENIX group is re-analyzed using the present models.  Lastly, we state our main findings in the Conclusion.

\section{Theoretical Background}

\subsection{One Body Density Matrix}

The formal definition of Bose condensation in a strongly interacting system is given in terms of the one-body density matrix $\widetilde{n}(s)$\cite{PSbook, Anderson}.  This quantity is the expectation value of the product of a particle creation operator at $\mathbf{s}$ and a particle annihilation operator at the origin: $\widetilde{n}(s)= \langle\Psi^\dagger(\mathbf{s})\Psi(0)\rangle$.  At zero temperature, $\widetilde{n}(s)$ measures the overlap in the many-body wavefunction of the system when a particle is removed from the origin and then placed some distance $s$ away.  A system is Bose-condensed if and only if $\widetilde{n}(s)$ approaches a constant value, $n_0$, as $s\rightarrow\infty$.  The atomic momentum distribution $n(k)$ is the Fourier transform of the one-body density matrix $\widetilde{n}(s)$.  Accordingly, the Bose condensate appears in $n(k)$ as a $\delta$-function located at $k = 0$ with an integrated intensity of $n_0$.

The one-body density matrix $\widetilde{n}(s)$ of an interacting Bose system as a function of temperature can be computed from first principles, i.e., directly from a microscopic Hamiltonian making use of a realistic potential to describe the interaction among atoms, by means of Quantum Monte Carlo simulations. In particular, the worm algorithm (WA) in continuous space\cite{worm,worm2}, based on Feynman's space-time formulation of quantum statistical mechanics\cite{Feynman}, has emerged over the past decade as a powerful methodology,  allowing one to calculate accurate equilibrium thermodynamics of Bose systems. The values of the physical estimates can be regarded as {\em exact}, at least for practical purposes, as the statistical and systematic uncertainties (the latter arising from the finite size of the simulated system) affecting them can be rendered in practice negligibly small with the typical computing resources available nowadays.

Because this numerical technique, which is by now fairly well established, is extensively described elsewhere\cite{worm, worm2}, we do not review it here. Rather, we merely furnish the relevant technical details. The results presented here were obtained by simulating an ensemble of $N=256$ $^4$He atoms enclosed in a cubic vessel, with periodic boundary conditions.  We used the standard microscopic model of $^4$He, based on the Aziz pair potential\cite{aziz79,aziznote}. In principle, a more complete microscopic description of the system should include contributions to the potential energy associated not just with pairs, but also with, e.g., triplets of atoms. Indeed, such contributions are known to play an important role when it comes to reproducing theoretically the experimental equation of state of liquid $^4$He, but their effect of the single-particle dynamics (e.g., the kinetic energy) has been shown to be relatively small\cite{moroni,pederiva}.  Thus, the neglect of three (and higher) body terms in the Hamiltonian, in a theoretical calculation aiming mainly at reproducing the value of the condensate fraction, is widely regarded as justified. 

Our simulations are carried out at fixed density, using a canonical variant of the WA\cite{mezzacapo,mezzacapo2}; the values of the density corresponding to the various temperatures (in the range $0.5 \le T\le 2.65$ K)  are taken from Ref. \onlinecite{barenghi}. We report results extrapolated to the limit of vanishing imaginary time step $\tau$ (see Ref. \onlinecite{worm2} for details). In general, results obtained with $\tau=1/640$ K$^{-1}$ are indistinguishable from the extrapolated ones, within the statistical uncertainty of the calculation.  We estimate the potential energy contribution arising from particles outside the main simulation cell by setting the pair correlation function $g(r)$ to unity outside the cell; which is an excellent approximation for the system size utilized in this work. The value of the ground state energy extrapolated to temperature $T=0$ is $-7.182\pm 0.013$ K per $^4$He atom, indistinguishable from the estimate at the lowest temperature considered here ($T=0.5$ K), within statistical uncertainties. 

\begin{figure}
\includegraphics[width=\columnwidth, keepaspectratio]{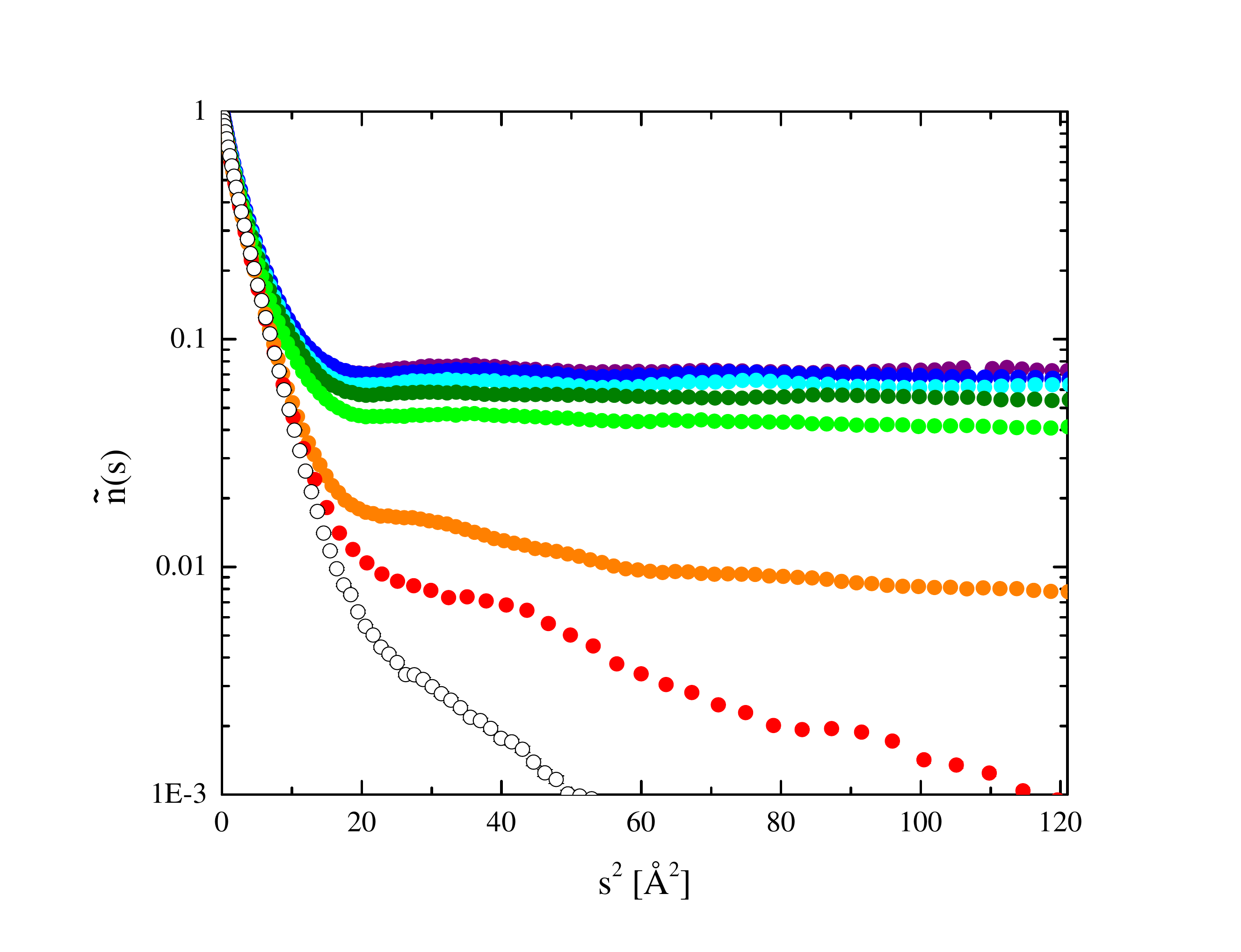}
\caption{QMC calculations of the one-body density matrix $\widetilde{n}(s)$ of liquid $^4$He under saturated vapor pressure: 1.09 K (purple), 1.40 K (blue), 1.65 K (cyan); 1.80 K (dark green), 2.00 K (light green), 2.35 K (orange), 2.65 K (red), 4.2 K (open circles).  Errors are smaller than symbol size.}
\label{fig:OBDM}
\end{figure}

Figure \ref{fig:OBDM} plots the calculated one-body density matrix $\widetilde{n}(s)$ for the conditions of our experiment.  Differences between the normal and superfluid phases are evident.  In the normal fluid, $\widetilde{n}(s)$ decays toward zero at large $s$; in the superfluid, $\widetilde{n}(s)$ reaches a constant value $n_0$ at large $s$.  The condensate fraction $n_0$ varies rapidly just below $T_\lambda$ and approaches a constant value around 1 K. The estimate at $T=1.09$ K is 0.075(2), which is consistent with that of Ref. \onlinecite{worm2}. The extrapolated $T=0$ value is 0.076(2), also consistent, within the quoted statistical uncertainties, with the estimate provided in Ref. \onlinecite{moroni}.

The average atomic kinetic energy $\langle E_K\rangle$ is given by the curvature of $\widetilde{n}(s)$ at $s = 0$.  Specifically, $\langle E_K\rangle = -(\hbar^2/2m)\nabla^2\widetilde{n}(s)$, the Laplacian being evaluated at $s = 0$.  Theoretical predictions for the average kinetic energy $\langle E_K\rangle$ and condensate fraction $n_0$ are given in Tables \ref{table:KE} and \ref{table:n0_ARCS} respectively.

\subsection{Neutron scattering and the Incoherent Approximation}

Here we review the theoretical basis for neutron Compton scattering studies of liquid $^4$He\cite{HohenbergPlatzman, SilverSokol, Glyde}.  One measures the double-differential scattering cross section in an inelastic neutron scattering experiment:  
\begin{equation} \label{eq:doubledif}
\frac{d^2\sigma}{d\Omega dE} = b_{\textrm{coh}}^2\frac{k_i}{k_f}S(Q, E).
\end{equation}
Here $b_{\textrm{coh}}$ is the coherent scattering length of $^4$He and $k_i$ ($k_f$) is the incident (final) neutron wavevector.  There is no incoherent contribution to the scattering from changes in the spin state of the $^4$He nucleus($b_{\textrm{inc}} = 0$).  The dynamic structure factor $S(Q, E)$ of a quantum liquid is the Fourier transform of its time-dependent density-density correlation function.
\begin{equation} \label{eq:SQE}
S(\mathbf{Q}, E) = \frac{1}{2\pi N}\int_{-\infty}^{+\infty}e^{iEt/\hbar}\langle\rho(\mathbf{Q}, t)\rho^\dagger(\mathbf{Q}, 0)\rangle dt.
\end{equation}
We may distinguish between two different regimes of $Q$ and $E$ transfer.  At low $Q (\lesssim 4 \textrm{ \AA}^{-1})$, the measured scattering is dominated by coherent interference between particles and hence the collective excitations (the phonon-roton modes) of the liquid are observed\cite{Andersen1, Andersen2}.  At high $Q (\gtrsim 10 \textrm{ \AA}^{-1})$ the coherent interference between different particles is cancelled by rapid phase variations.  The scattering is now dominated by single particle excitations.  Therefore, the incoherent approximation is used to reduce the dynamic structure factor to
\begin{equation}
S^{(i)}(Q, E) = \frac{1}{2\pi}\int_{-\infty}^{+\infty} e^{iEt/\hbar}\langle e^{-\mathbf{Q}\cdot\mathbf{r}(t)}e^{i\mathbf{Q}\cdot\mathbf{r}(0)} \rangle dt.
\end{equation}
At high $Q$, the central moments of the scattering obey the following sum rules\cite{Placzek}:
\begin{eqnarray}
\textrm{Normalization:} \int_{-\infty}^{+\infty} S(Q, E) dE &=& 1.\\
\textrm{$f$-sum rule:} \int_{-\infty}^{+\infty} (E - E_R) S(Q, E) dE &=& 0.\\
\textrm{$\omega^2$-sum rule:} \int_{-\infty}^{+\infty} (E - E_R)^2S(Q, E) dE &=& \frac{4}{3}E_R\langle E_K\rangle.
\end{eqnarray}
Here $E_R = \hbar^2Q^2/2m$ is the recoil energy of a helium atom.

\subsection{Impulse Approximation}

The Impulse Approximation (IA) assumes that the kinetic energy transferred by an incident neutron to an individual helium atom during a scattering event is so large that the potential energy of the atom in both its initial and final states may be neglected.  The IA is valid at infinite $Q$ so long as the interatomic potential does not contain a hard core.  Within the IA, $S^{(i)}(Q, E)$ reduces to an integral transform of the momentum distribution $n(k)$:
\begin{equation}\label{eq:IA}
S_{\textrm{IA}}(Q, E) = \int n(\mathbf{k})\delta\left(E - \frac{\hbar^2Q^2}{2m} - \frac{\mathbf{k}\cdot\mathbf{Q}}{m}\right)  d\mathbf{k}.
\end{equation}
Here $n(\mathbf{k})$ is the atomic momentum distribution.  When the IA is satisfied, a constant $Q$ cut of the dynamic structure factor $S_{\textrm{IA}}(Q, E)$ consists of a single peak symmetric about the recoil energy $E_R = \hbar^2Q^2/2m$.  The width of the peak is proportional to the product of $Q$ and the width of $n(k)$.  The sum rules of incoherent scattering also apply at infinite $Q$, as the IA is a special case of the incoherent approximation.  

Typically, the scattering data is presented and analyzed in terms of the West scaling variable $Y$ and the neutron Compton profile $J(Y, Q)$\cite{Sears}.  These quantities are defined as follows:
\begin{subequations}
\begin{equation}
Y = \frac{m}{\hbar^2 Q}\left(E - \frac{\hbar^2 Q^2}{2m}\right).
\end{equation}
\begin{equation}
J(Y, Q) = \frac{\hbar^2 Q}{m}S(Q, E).
\end{equation}
\end{subequations}

If the IA is valid, then the neutron Compton profile $J_{IA}(Y)$ is related the atomic momentum distribution $n(\mathbf{k})$ by a Radon transform\cite{Andreani}.  The atomic momentum distribution of a Bose-condensed fluid may be expressed as a sum: $n(\mathbf{k}) = n_0\delta(\mathbf{k}) + n^*(\mathbf{k})$, where the $\delta$-function singularity is due to the condensate.  Expressing \ref{eq:IA} in terms of the scaling variable $Y$ yields:
\begin{equation}
J_{IA}(Y) = n_0\delta(Y) + 2\pi\int_{|Y|}^\infty kn^*(k)dk.
\label{eq:JIA}
\end{equation}

There are several advantages to analyzing the scattering data in terms of the scaling variable $Y$.  First, the neutron Compton profile $J(Y, Q)$ is a one-dimensional projection of the momentum distribution $n(\mathbf{k})$.  In the IA, the $Y$-scaling variable has the physical interpretation of being the component $k_{\parallel}$ of the atomic momentum that is parallel to the momentum transfer $\mathbf{Q}$ from the incident neutron $k_{\parallel} = \mathbf{k}\cdot\mathbf{\hat{Q}}$.  The West scaling variable $Y$ is also the Fourier conjugate of the distance $s$ traveled by a recoiling helium atom.  Second, $J(Y, Q)$ scales with $Q$.  Such behavior is necessary, but not sufficient, to demonstrate the applicability of the IA.

The scattering in the IA-limit is obtained from the one-body density matrix $\widetilde{n}(s)$ by a Fourier cosine transform.  Figure \ref{fig:IAscatt} compares $J_{\textrm{IA}}(Y)$ at 1.09 K and 2.65 K.  The most striking feature of $J_{\textrm{IA}}(Y)$ in the superfluid phase is the $\delta$-function singularity at $Y = 0$.

\subsection{Final State Effects}

The straightforward predictions of the IA turn on the assumption that a target helium atom recoils freely from the impact of a high energy neutron.  However, the interatomic potential has a steeply repulsive core at short distances, making interactions of the recoiling atom with its neighbors important even at high $Q$.  The resulting deviations from the IA are known as Final State Effects (FSE).  Hohenberg and Platzman argued that the FSE broadening is governed by the $^4$He-$^4$He scattering cross section $\sigma(Q)$\cite{HohenbergPlatzman,Feltgen}.  They estimated that the condensate peak would be broadened by an amount roughly equal to $\rho\sigma(Q)$, where $\rho$ is the number density of the liquid.  For $Q = 30 \textrm{ \AA}^{-1}$, the broadening is on the order of $0.7 \textrm{ \AA}^{-1}$, which is not small compared to the expected width of $J_{\textrm{IA}}(Y)$, namely $\approx2 \textrm{ \AA}^{-1}$.

Several theoretical approaches to understanding final state effects have been proposed. In general, these approaches fall into one of three categories.  The first treat final state effects as a convolution with the IA scattering.  This approach has been followed by Gersch and Rodriguez\cite{GerschRodriguez,Mazzanti}, Silver\cite{Silver0,Silver1,Silver2}, Carraro and Koonin\cite{CK,Moroni2} and Glyde\cite{Glyde}.  A second approach, which has been used by Sears\cite{Sears}, treats final state effects as a additive correction to the IA scattering.  Finally, there are theories that treat final state effects by other methods such as alternate scaling variables\cite{Stringari}.   A detailed comparison of these theories is beyond the scope of this work. 

We will focus on the theories that treat final state effects in terms of a broadening function:
\begin{equation}\label{eq:JFS}
J_{FS}(Y, Q) = \int_{-\infty}^{+\infty} J_{IA}(Y^\prime)R(Y - Y^\prime, Q)dY^\prime.
\end{equation}
where $R(Y,Q)$ is the final state broadening function.  Such theories can be separated into two classes: those that calculate $R(Y,Q)$ \emph{a priori} from known quantities, such as the interatomic potential and pair distribution function, and those where the parameters of $R(Y,Q)$ must be obtained from the scattering.  We examine the theories of Silver and Carraro and Koonin since they provide concrete, testable predictions for the form of $R(Y, Q)$.

The sum rules for incoherent scattering place constraints on the neutron Compton profile $J(Y, Q)$ and the FSE function $R(Y, Q)$.  These rules require that $J(Y, Q)$ and $R(Y, Q)$ both be normalized to unity and have a zero first moment.   The second moment of these functions must satisfy:
\begin{subequations}
\begin{equation}
\int_{-\infty}^{+\infty} Y^2 J(Y, Q) dY =\frac{2m}{3\hbar^2} \langle E_K \rangle.
\end{equation}
\begin{equation}
\int_{-\infty}^{+\infty} Y^2 R(Y, Q) dY = 0.
\end{equation}
\end{subequations}
The $\omega^2$-sum rule implies that the second moment of $R(Y, Q)$ is identically zero.  This means that $R(Y, Q)$ cannot be represented by a simple, positive-definite function, such as a Gaussian or a Lorentzian.  Instead, $R(Y, Q)$ must assume both positive and negative values.  The effect of convoluting $J_{\textrm{IA}}(Y)$ with $R(Y, Q)$ is not only to broaden the condensate peak, but also to redistribute intensity around the spectrum so that the second moment of the scattering is unaffected.

Silver developed a model lineshape $R_{\textrm{S}}(Y, Q)$ for the FSE corrections in liquid $^4$He using Hard Core Pertubation Theory\cite{Silver1, Silver2, SilverSokol}.  The theory takes the interatomic potential and pair-distribution function as inputs.  An intuitive picture underlies Silver's theory.  Before the scattering event, each helium atom is located near the minimum of the potential well generated by its nearest neighbors and far from the repulsive cores responsible for final state effects.  During the impact of a high energy incident neutron, the recoiling helium atom travels a distance $s$, over which it may encounter the steeply repulsive cores of its neighbors.  On this theory, the scaling variable $Y$ is conjugate to the recoil distance $s$, although $Y$ is no longer identical to $k_{\parallel}$.  The FSE broadening function $R_{\textrm{S}}(Y, Q)$ is related to the Fourier transform of the classical scattering probability of suffering no collisions as a function of the travel distance $s$.  

Carraro and Koonin developed an alternative theory $R_{\textrm{CK}}(Y, Q)$ for FSE corrections\cite{CK}.  The starting point of their calculation is a Jastrow approximation to the many-body wavefunction of liquid $^4$He.  They calculate the propagator for a single atom moving at a high $Q$ within the static potential generated by the instantaneous configuration of background atoms, and the result is averaged over many configurations distributed according to the variational wavefunction.  As in Silver's model, the scaling variable $Y$ has the physical interpretation of being the Fourier conjugate variable to the travel distance $s$.  Here we use an improved scheme whereby the background atoms are distributed according to a better approximation to the exact ground state than afforded by a Jastrow wavefunction.  Details are given in Ref \onlinecite{Moroni2}.  We have calculated $R_{\textrm{S}}(Y, Q)$ and $R_{\textrm{CK}}(Y, Q)$ using the Aziz potential\cite{aziz79}.

Both the Silver and Carraro-Koonin theories of FSE are designed for the ground state.  We assume that the temperature dependence of the Compton profile $J_{\textrm{FS}}(Y, Q)$ in Equation \ref{eq:JFS} is restricted to the factor $J_{\textrm{IA}}(Y^\prime)$, through the one-body density matrix $\widetilde{n}(s)$.

\begin{figure}
\includegraphics[width=0.9\columnwidth, keepaspectratio]{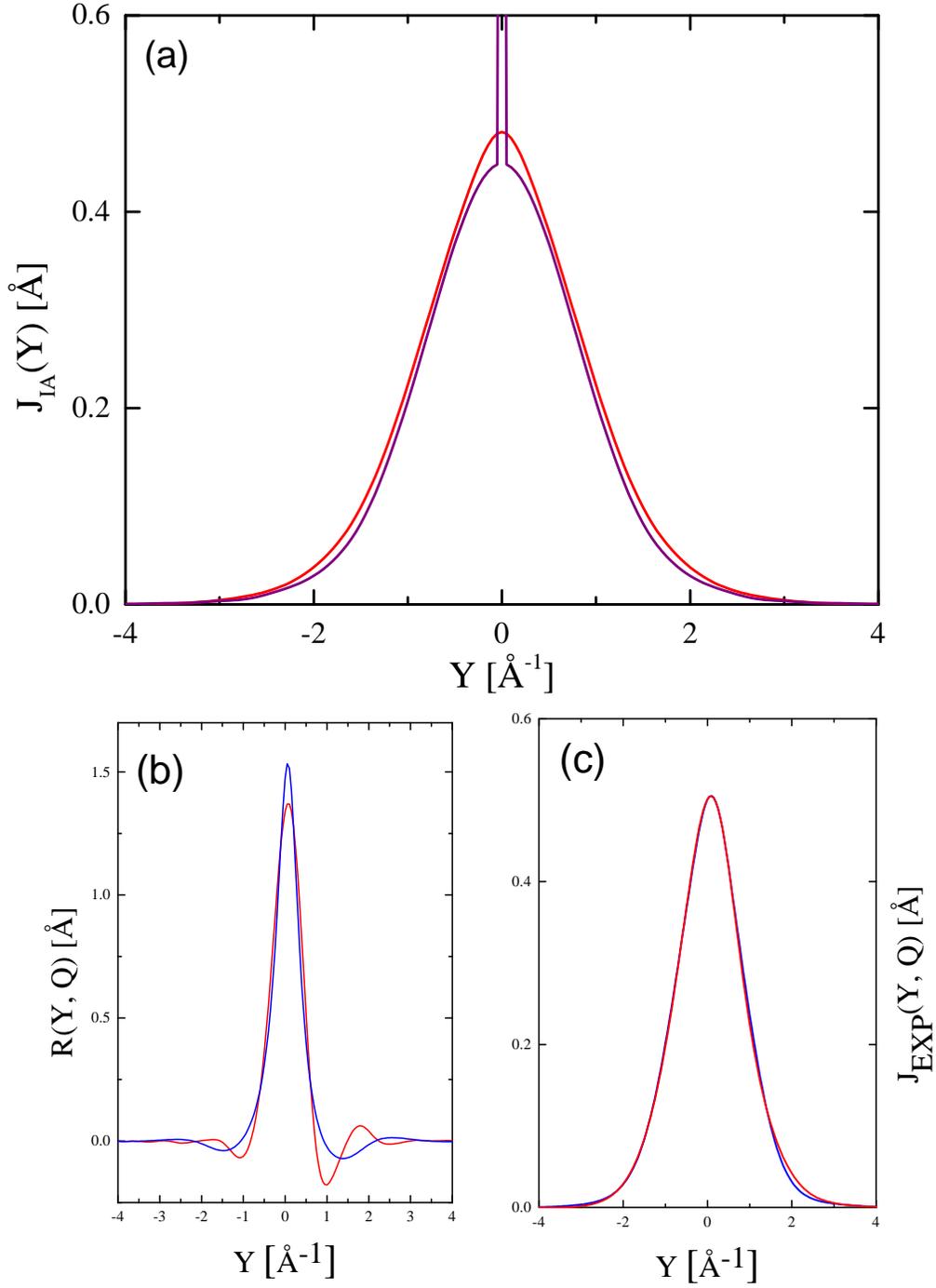}
\caption{Panel (a): Expected IA-scattering $J_{\textrm{IA}}(Y)$ at temperatures of 2.65 K and 1.09 K.  Panel (b): Comparison of final state effect functions $R(Y, Q)$ at $Q = 27.0 \textrm{ \AA}^{-1}$ and a liquid number density $\rho = 0.02187 \textrm{ \AA}^{-3}$.  Curves: Hard Core Perturbation Theory (blue); Carraro-Koonin theory (red).  Panel (c): The expected scattering $J_{\textrm{EXP}}(Y, Q)$ at $T = 1.09 \textrm{ K}$ when the IA is broadened by final state effects and instrumental resolution.}
\label{fig:IAscatt}
\end{figure}

Figure \ref{fig:IAscatt}(b) compares the predictions of the Silver and Carraro-Koonin theories at a wavevector $Q = 27.0 \textrm{ \AA}^{-1}$ and the equilibrium number density $\rho = 0.0217 \textrm{ \AA}^{-3}$.  Both models consist of a central peak and damped oscillatory tails which are both positive and negative.  They obey the normalization, $f$-sum rule, and $\omega^2$-sum rule conditions.  The central peak of $R_{\textrm{CK}}(Y, Q)$ (FWHM $\approx 0.8 \textrm{ \AA}^{-1}$) is broader than that of $R_{\textrm{S}}(Y, Q)$ (FWHM $\approx 0.6 \textrm{ \AA}^{-1}$).  The oscillatory tails of the two theories are out of phase, and they have different frequencies and amplitudes.

The expected intrinsic scattering $J_{\textrm{FS}}(Y, Q)$  is obtained by convoluting the QMC calculations of $J_{\textrm{IA}}(Y)$ with the FSE broadening functions.  Figure \ref{fig:IAscatt}(c) illustrates the anticipated scattering, including the effects of instrumental resolution, at $T = 1.09 \textrm{ K}$ and $Q = 27.0 \textrm{ \AA}^{-1}$.  Despite obvious differences between the Silver and Carraro-Koonin theories, their predictions are similar.  The lineshape is broad and featureless: the condensate peak has entirely disappeared.  Small differences in the predicted lineshape will be undetectable in the presence of statistical noise.  The only practically observable difference between the Silver and Carraro-Koonin theories occurs near $Y = +2 \textrm{ \AA}^{-1}$.

Finally, we note the IA is approached slowly as a function of $Q$.  If the interatomic potential had an infinitely hard core, then $\sigma(Q)$ would be independent of $Q$.  The neutron Compton profile $J(Y, Q)$ would $Y$-scale even though the IA limit would never be reached\cite{Weinstein}.  However, the `real' interatomic potential is exponentially repulsive at short distances.  Accordingly, the $^4$He-$^4$He scattering cross section $\sigma(Q)$ varies as $\log{Q}$, apart from glory oscillations\cite{Feltgen}.  Therefore, $Y$-scaling should hold, to good approximation, over a limited range in $Q$, although the scaling function will not be $J_{\textrm{IA}}(Y)$.

\section{Experimental Approach}

\subsection{Experimental Details}

We carried out a neutron Compton scattering study of liquid $^4$He using the ARCS spectrometer\cite{ARCS, Stone, AbernathyMonitor} at the Spallation Neutron Source.  This instrument is a direct geometry, time-of-flight spectrometer.  Incident neutron energies between 15-5000 meV are available from the decoupled poisoned water moderator.  A $T_0$ chopper, operating at a frequency of 180 Hz, blocks the burst of prompt radiation released from the source when the protons hit the target.  An incident neutron energy ($E_i = 710 \textrm{ meV}$) is chosen by time-of-flight using the phase of a Fermi chopper rotating at 600 Hz, placed just upstream of the sample.  The sample was enclosed in a cryostat which will be described shortly.  Neutrons that scatter off of the sample traverse an oscillating radial collimator on their way to the detector array.  There are two low efficiency beam monitors, one located after the Fermi chopper and another located just before the beam stop.  The beam profile observed at these monitors is used to determine the initial neutron energy $E_i$ and moderator emission time $t_0$.  Complete details of the instrument are provided in Refs \onlinecite{ARCS, Stone}.

The aforementioned sample environment consisted of an orange cryostat coupled with a custom 1 K insert.  The orange cryostat cooled the insert to a temperature $< 3$ K.  The custom insert then provides a base temperatures of approximately 1 K and temperature stability of $< 1$ mK. It consists of  Al-6061 sample cell that is mechanically mounted to a 1 K pot built from oxygen-free high conductivity copper.  Both are enclosed within an aluminum vacuum can isolating them from the exchange gas of the orange cryostat.  The liquid $^4$He within the sample cell had a height of 5.08 cm and a diameter of 2.54 cm.  We estimate that the beam transmission is approximately 94\% given the sample geometry.  Two temperature control packages, consisting of a heater and a germanium semiconductor thermometer, were attached to the insert, one to the 1 K pot and the other to the bottom of the sample cell.  The temperature stability was obtained by operating each package on an independent temperature control loop.  Each thermometer was calibrated to $\pm$ 4 mK.  Furthermore, the temperature dependence of the observed vapor pressure of the liquid $^4$He in the sample cell was consistent with the semiconductor thermometry.

We collected data at a series of different sample temperatures $T = $ 1.090(10), 1.400(10), 1.650(4), 1.800(4), 2.000(4), 2.100(4), 2.35(30), and 2.650(15) K with measurement periods varying between 2 hours and 15 hours.  The background scattering due to the sample environment and empty sample can was measured at 2.7(1) K.   The quoted errors in the sample temperature represent whichever quantity is larger, either the systematic uncertainty in the temperature scale or the random uncertainty from the stability of the cryogenics.  No thermal gradient was observed between the bottom of the sample cell and the 1 K pot above it.    

An event-based data acquisition system stores the data as list of time stamps and pixel locations.  Histogramming the raw data occurs during reduction and at this step  we filtered out events that occurred when the sample temperature was outside of our stability criteria\cite{peterson2015event}.  The data, as counts versus time-of-flight, was then normalized to the proton charge on target to remove variation in source output.  The measured double-differential cross section $d^2\sigma/d\Omega dE$ is transformed to the neutron Compton profile $J(Y, Q)$ using the Mantid and DAVE software packages.\cite{Mantid, DAVE}  The scattering data $J(Y, Q)$ between $20.0 \textrm{ \AA}^{-1}\le Q \le 27.5 \textrm{ \AA}^{-1}$ was analyzed in steps of $0.5 \textrm{ \AA}^{-1}$ each having a widths of $\pm 0.2 \textrm{ \AA}^{-1}$.

We used two independent methods to determine the absolute intensity scale for the neutron Compton profile $J(Y, Q)$.   One approach is to measure the total scattering off a standard vandadium foil having the same dimensions as the lateral surface area of the sample can, and scaling the observed double-differential cross section $d^2\sigma/d\Omega dE$ of the sample accordingly.    Taking the microscopic scattering cross section of vanadium to be 421.0 mbarn/sR, the normalization factor was determined by integrating the scattering over energy transfers $-150 \textrm{ meV} \leq E \leq 685 \textrm{ meV}$ and scattering angles $10^\circ \le \phi \le 135^\circ$.  In the second approach, we numerically integrated the neutron Compton profile $J(Y, Q)$ and imposed the zeroth moment sum rule.  This method of setting an absolute intensity scale implicitly assumes that all of the nonzero parts of $J(Y, Q)$ are observed.  The two methods typically produced consistent absolute intensity scales in the range of $1\%$ to $10\%$. 

When comparing the experimental data to theoretical models, we allow for a small shift $Y_c$ in the theoretical predictions to ensure that the $f$-moment sum rule is satisfied.  If the energy scale were perfectly defined then $Y_c$ would be exactly zero. However, uncertainties in the incident neutron energy $E_i$, moderator emission time $t_0$, lengths of flight paths, and other instrument parameters can introduce small shifts to the energy scale.  Typical values for $Y_c$ are on the order of $0.01 \textrm{ \AA}^{-1}$ which is less than a bin width in $Y$ and small compared to the resolution width.

\subsection{Instrumental Resolution}

We calculated the instrumental resolution function $I(Y, Q)$ using a realistic Monte Carlo ray tracing simulation of the scattering experiment\cite{ARCSsim, ARCS_sim}.  The simulations were carried out using the McStas software suite\cite{McStas1, McStas2}.  The input to the simulation includes the spectrum of the decoupled water moderator, the known instrument parameters, sample geometry, and a sample kernel.  There were $2.8\times 10^{12}$ incident neutron pulses simulated for this calculation.  The output of the simulation includes both the incident beam monitors and the scattering measured at the detector bank.  The output of the simulation, as counts versus time-of-flight, receives the same treatment as the real scattering data.  The effective instrumental resolution function $I(Y, Q)$ is determined from the output of the simulation by deconvoluting the known sample kernel from the simulated scattering.

An accurate description of the time-structure of the incident neutron pulse is necessary for a reliable determination of $I(Y, Q)$.  We found that the McStas model reproduces the time-of-flight profiles observed by the incident beam monitors.  This indicates that the instrument simulation faithfully describes the time-distribution of neutrons as they emerge from the moderator and pass through the Fermi chopper.

Recent calculations of the instrumental resolution function $I(Y, Q)$ of ARCS used an ideal $\delta$-function scatterer for the sample kernel\cite{Diallo201634}.  Instead, we have chosen to base our sample kernels on previous Quantum Monte Carlo calculations of $n(k)$\cite{GFMC, CepReview}.  The  istropicSqw sample component was used as it allowed us to easily change models\cite{Farhi20095251}.  

We found that the effective resolution function $I(Y, Q)$ could be described as a single Gaussian in $Y$.  The full-width at half-maximum of $I(Y, Q)$ decreases roughly linearly from $1.05 \textrm{ \AA}^{-1}$ at $Q = 20 \textrm{ \AA}^{-1}$ to $0.50 \textrm{ \AA}^{-1}$ at $Q = 27.5 \textrm{ \AA}^{-1}$.  Our calculated resolution functions $I(Y, Q)$ agree with the `observed' ones reported in Ref. \onlinecite{Diallo201634}.

The observed Compton profile $J_{\textrm{EXP}}(Y, Q)$ is obtained by convoluting the intrinsic scattering with the instrumental resolution function:
\begin{equation}
J_{\textrm{EXP}}(Y, Q) = \int_{-\infty}^{+\infty} J_{\textrm{FS}}(Y^\prime, Q)I(Y - Y^\prime, Q)dY^\prime
\end{equation}

Both the instrumental resolution $I(Y, Q)$ and FSE function $R(Y, Q)$ have the effect of smearing sharp features in $J_{\textrm{IA}}(Y)$.

\subsection{Background Subtraction and Multiple Scattering Corrections}

The sample-independent background scattering was measured at $2.7(1) \textrm{ K}$.  The signal is due to scattering from the sample cell, insert vacuum can, the tails of the orange cryostat, and dark counts.  At the wavevectors considered in the data analysis, $20.0 \textrm{ \AA}^{-1} \leq Q \leq 27.5 \textrm{ \AA}^{-1}$, the helium recoil peak is either mostly or completely separated from the elastic Bragg scattering and heavy element recoil lines present in the background.  The signal-to-background ratio in the region of the helium peak is very high.

We find that a sample-dependent residue remains after the subtraction of the background signal.  This component of the measured signal is due to the multiple scattering of neutrons.  It is approximately constant with scattering angle and forms a broad band in the energy spectrum, being centered at 300 meV and having a FWHM of 380 meV.  The intensity of the multiple scattering is only a few percent of the intensity of the helium peak.

Here we make the assumption that the multiple scattering is isotropic and additive.  Because the multiple scattering at low $Q$ is clearly separated from the helium recoil line, we fit the multiple scattering component at low $Q$ to a smooth curve and subtracted this smooth curve from the experimental data at all values of $Q$.

\section{Experimental Results}

\begin{figure}
\includegraphics[width=0.8\textwidth, height=0.8\textheight, keepaspectratio]{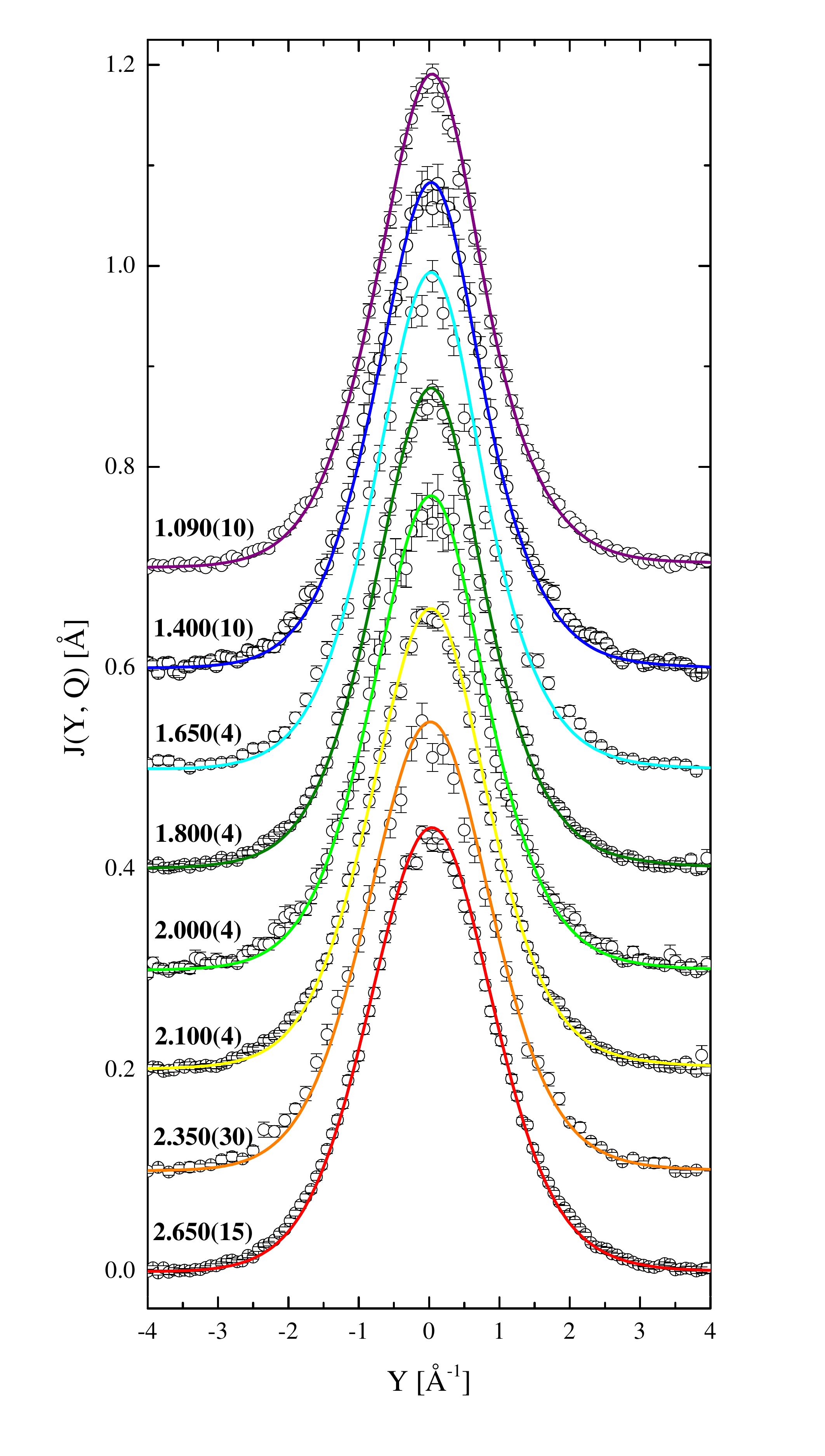}
\caption{The neutron Compton profile $J(Y, Q)$ at $Q = 27 \textrm{ \AA}^{-1}$.  The different temperature data sets have been vertically offset by $0.1 \textrm{ \AA}^{-1}$.  The solid lines represent our QMC calculations folded with the instrumental resolution function $I(Y, Q)$ the final state effect function $R(Y, Q)$ of Carraro-Koonin.  Throughout the paper, error bars on the scattering data represent one standard deviation.
\label{fig:waterfall}}
\end{figure}

Figure \ref{fig:waterfall} plots the neutron Compton profile $J(Y, Q)$ observed at $Q = 27.0 \textrm{ \AA}^{-1}$ as a function of temperature.  The observed scattering $J(Y, Q)$ consists of a single, non-Gaussian peak containing no sharp features or oscillations.  The overall width of the scattering $\approx 2 \textrm{ \AA}^{-1}$ is dominated by quantum-mechanical zero-point motion.  It is also much broader than the instrumental resolution width, $0.55 \textrm{ \AA}^{-1}$, at this $Q$.

We find that the scattering is only weakly dependent on temperature in the normal fluid phase.  When the temperature is reduced below $T_\lambda$, the scattering $J(Y, Q)$ becomes visibly narrower and more peaked.  This increase in scattering at small $Y$ below $T_\lambda$ is consistent with the existence of a Bose condensate peak located at $Y = 0$ which has been broadened by finite instrumental resolution and FSE.  However, as shown below, the scattering data is also consistent with models that do not include a Bose condensate.  The scattering data does not, by itself, prove that a Bose broken symmetry is responsible for the phase transition at $T_\lambda$.
 
\begin{figure}
\includegraphics[width=\textwidth, height=\textheight, keepaspectratio]{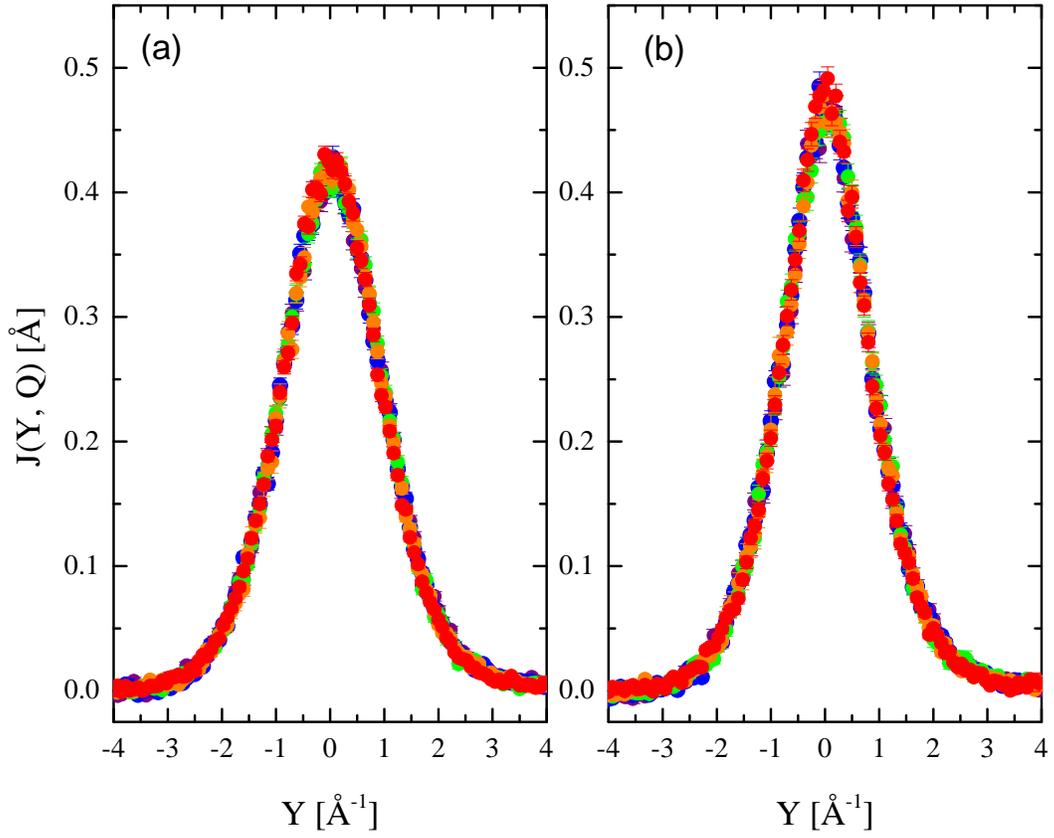}
\caption{Test of Y-scaling at (a) 2.650(15) K and (b) 1.090(10) K.  Points: $Q = 23.0 \textrm{ \AA}^{-1}\textrm{(purple)}, 24.0 \textrm{ \AA}^{-1}\textrm{(blue)}, 25.0 \textrm{ \AA}^{-1}\textrm{(green)}, 26.0 \textrm{ \AA}^{-1}\textrm{(orange)}, 27.0 \textrm{ \AA}^{-1}\textrm{(red)}$.
\label{fig:Yscale}}
\end{figure}

$Y$-scaling behavior is observed at all temperatures considered in this study.  To illustrate, Figure \ref{fig:Yscale} overplots the scattering in the normal and superfluid phases.  In both cases, the scattering clearly collapses onto a single curve.

One might be tempted to conclude from this fact that the IA-regime has been reached in this experiment.  We stress that $Y$-scaling is a necessary, but not sufficient, condition for the IA.  Because FSE in liquid $^4$He vary as $\log(Q)$, they are expected to not change appreciably over less than a decade in $Q$.  As a result, the scattering data obeys $Y$-scaling to good approximation, even though the scaling function is not $J_{\textrm{IA}}(Y)$.

\section{Discussion}

\subsection{Lineshape Comparison}\label{sub:lineshape}

Theoretical calculations of the momentum distribution $n(k)$ may be checked for their consistency with the scattering data, even if the Bose condensate peak does not appear as a distinct feature in $J(Y, Q)$.  To make the most stringent possible test, one should compare the entire predicted lineshape for $J(Y, Q)$ with the neutron Compton scattering data.  The solid lines in Figure \ref{fig:waterfall} are obtained when our QMC calculations of $J_{\textrm{IA}}(Y)$ are convoluted with final state effects $R_{\textrm{CK}}(Y, Q)$ and instrumental resolution $I(Y, Q)$.  We have allowed the amplitude and center position $Y_c$ of the predicted scattering to vary, but not the shape of the peak.  As can be seen, there is excellent agreement between the predicted and observed lineshapes at all temperatures.  The same level of agreement is obtained at other values of $Q$.  This convergence between \emph{ab initio} predictions and the measured scattering is strong evidence that a Bose broken symmetry is responsible for the superfluid phase transition of liquid $^4$He.  

In making this comparison, we are testing the combination of the QMC calculations and the Carraro-Koonin theory.  The scattering data, when corrected for instrumental resolution, only provides information about $J_{\textrm{FS}}(Y, Q)$.  It does not provide information about the IA-scattering $J_{\textrm{IA}}(Y)$ or FSE function $R(Y, Q)$ considered separately.  If one assumes that the FSE function $R(Y, Q)$ is known, then one may test theoretical predictions for the IA-scattering $J_{\textrm{IA}}(Y)$ against the data.  Below we will introduce parameterized models for  $J_{\textrm{IA}}(Y)$ which permit empirical estimates for the average kinetic energy $\langle E_K\rangle$ and Bose condensate fraction $n_0$.  

\begin{figure}
\includegraphics[width=\textwidth, height=\textheight, keepaspectratio]{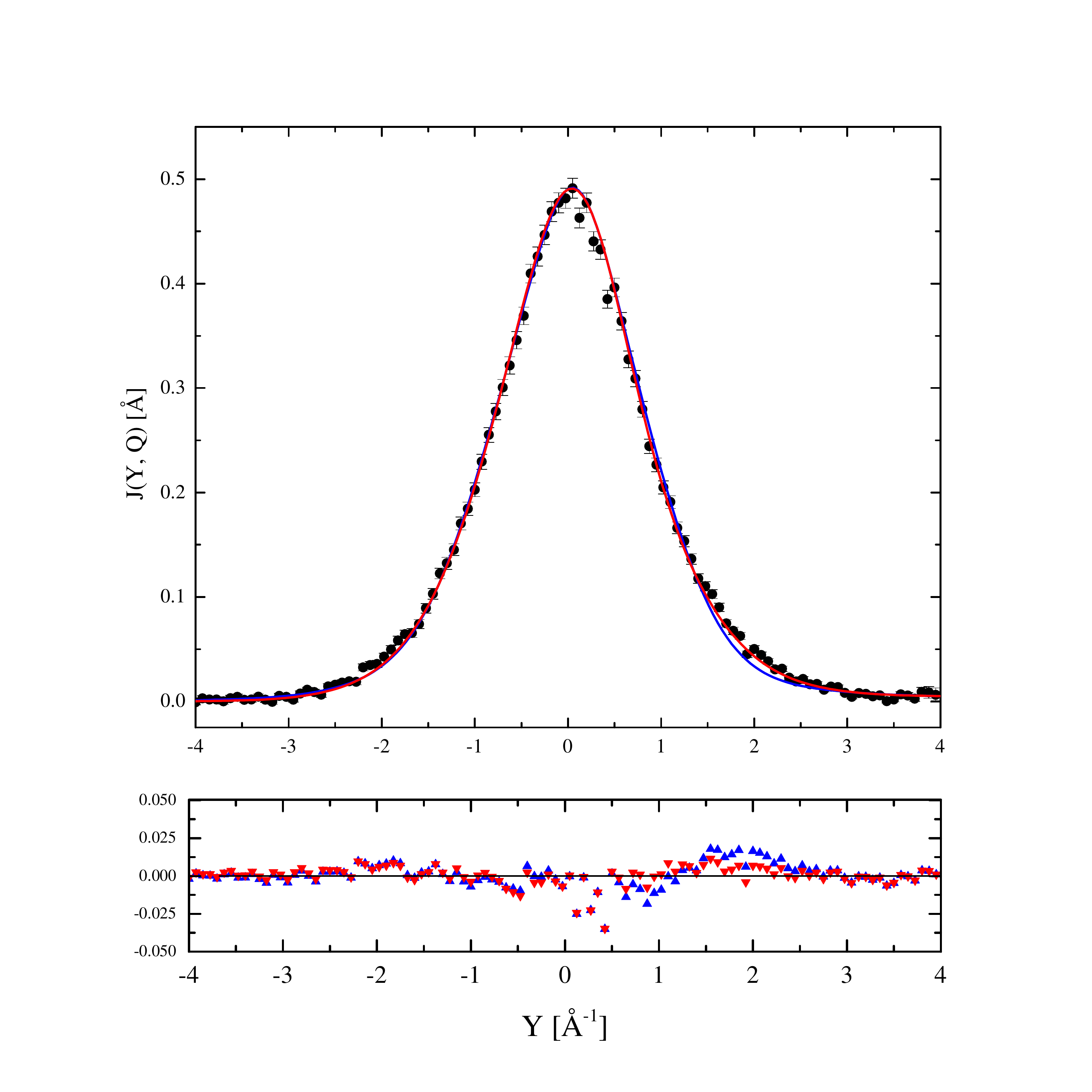}
\caption{The neutron Compton profile $J(Y, Q)$ at $Q = 27.0 \textrm{ \AA}^{-1}$ and $T = 1.090(10) \textrm{ K}$ is compared to theoretical predictions based on two different final state effect theories.  Main panel: experimental points (black circles); QMC calculations folded with Carraro-Koonin theory (red) and with HCPT (blue)  Lower panel: difference curves.
\label{fig:FSEcomp}}
\end{figure}

On the other hand, one may turn this problem around, assuming that $J_{\textrm{IA}}(Y)$ is known and test different theories for $R(Y, Q)$ against the scattering data.  Now we assume that our QMC calculations of $J_{\textrm{IA}}(Y)$ are correct.  The solid lines in Figure \ref{fig:FSEcomp} compared the predicted scattering according to the Silver and Carraro-Koonin theories with the experimental data at $T = 1.09 \textrm{ K}$ and $Q = 27.0 \textrm{ \AA}^{-1}$.  Overall, both theories are in excellent agreement with the scattering data.  Statistical noise and instrumental resolution effects make these theories indistinguishable for most values of $Y$.

Nevertheless, it is clear from the residuals shown in Figure \ref{fig:FSEcomp} that the Carraro-Koonin theory offers a better description of the scattering data near $Y = +2 \textrm{ \AA}^{-1}$.  We find that Silver's theory underestimates the scattering near $Y = +2 \textrm{ \AA}^{-1}$ at other temperatures $T$ and values of $Q$ as well.  For comparison, we note that the PHOENIX group adopted Silver's model FSE function in their comprehensive study of the $^4$He phase diagram\cite{Snow, SnowLong}.  Those authors also found a small systematic deviation near $Y = +2 \textrm{ \AA}^{-1}$ when $R_{\textrm{S}}(Y, Q)$ was used to analyze the scattering data.  Because the deviation was found not to depend upon the phase, temperature, density, or geometry of the sample, they attributed the small difference to the form of their model $R_{\textrm{S}}(Y, Q)$\cite{SnowLong}. 

\subsection{Empirical Estimates of $\langle E_K\rangle$}

Another approach to analyzing the scattering data is to define a parameterized model for the momentum distribution $n(k)$.  The values of the adjustable parameters are estimated by means of a least-squares fit to the experimental data, taking into account the broadening of the IA-scattering by instrumental resolution $I(Y, Q)$ and final state effects $R(Y, Q)$.  One thereby extracts several parameters from the experimental data, such as the average kinetic energy $\langle E_K\rangle$ or the Bose condensate fraction $n_0$, to compare to theoretical predictions.

We first employ the phenomenological model developed by Sosnick \emph{et al}\cite{SosnickLong, SnowLong} to obtain empirical estimates for $\langle E_K\rangle$ as a function of temperature.  Their model momentum distribution $n(k)$ consists of a sum of two Gaussians:
\begin{equation}
n_P(k) = \sum_{i = 1}^{2} \frac{A_i^\prime}{(2\pi\sigma_i^2)^{3/2}}e^{-k^2/2\sigma_i^2}
\end{equation}
where the integrated intensities, $A_1^\prime$ and $A_2^\prime$, add to unity.  This is a physically reasonable model for a cold quantum liquid where both particle statistics and zero-point motion are important.  It satisfies physical constraints such as being normalized, positive-definite, isotropic, and symmetric about $k = 0$.  The IA-scattering in this model is also given by a sum of two Gaussians:
\begin{equation} \label{Pmodel}
J_{\textrm{IA}}^{(P)}(Y) = \sum_{i = 1}^{2} \frac{A_i}{(2\pi\sigma_i^2)^{1/2}}e^{-(Y - Y_c)^2/2\sigma_i^2} + aY + b
\end{equation}
The two Gaussians are locked to a common center $Y_c$.  We have included the linear background in order to account for any multiple scattering that is not fully removed by our subtraction procedure.  The average kinetic energy is given by: $\langle E_K\rangle =  (3\hbar^2/2m)(A_1\sigma_1^2 + A_2\sigma_2^2)/(A_1 + A_2)$. 

\begin{figure}
\includegraphics[width=\textwidth, height=\textheight, keepaspectratio]{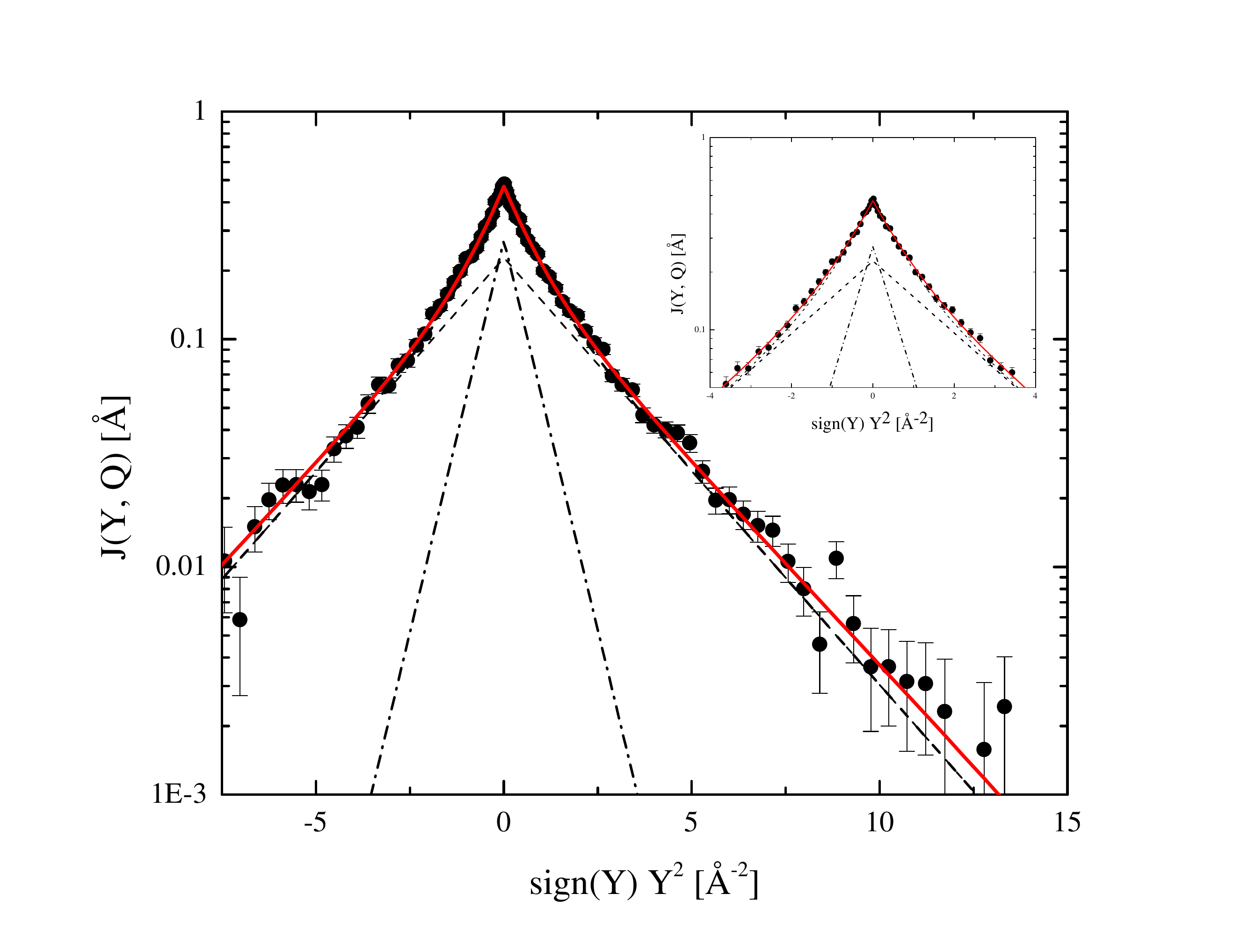}
\caption{The observed scattering at $Q = 26.5 \textrm{ \AA}^{-1}$ and $T = 1.800(4) \textrm{ K}$.  The blue curve is the result of fitting to the phenemonological model $J_{IA}(Y)$ as described in the main text.  The dashed and dash-dot lines are the two Gaussian components.  The value of $\chi^2$ is 1.006.
\label{fig:2G}}
\end{figure}

The observed neutron Compton profile $J(Y, Q)$ was fit using the phenomenological model $J_{\textrm{IA}}^{(P)}(Y)$ at all wavevectors $Q$ and temperatures $T$.  Figure \ref{fig:2G} plots a representative fit to the scattering data at $Q = 26.5 \textrm{ \AA}^{-1}$ and $T = 1.800(4) \textrm{ K}$.  The scattering data has been plotted as $\log\left(J(Y, Q)\right)$ vs $Y^2$ to illustrate each Gaussian component in the fit.  The small linear background due to multiple scattering has been subtracted.  Typical values of $\chi^2$ are close to one and the difference curves reveal no systematic discrepancies between the model and the scattering data.

\begin{figure}
\includegraphics[width=0.9\textwidth, height=0.9\textheight, keepaspectratio]{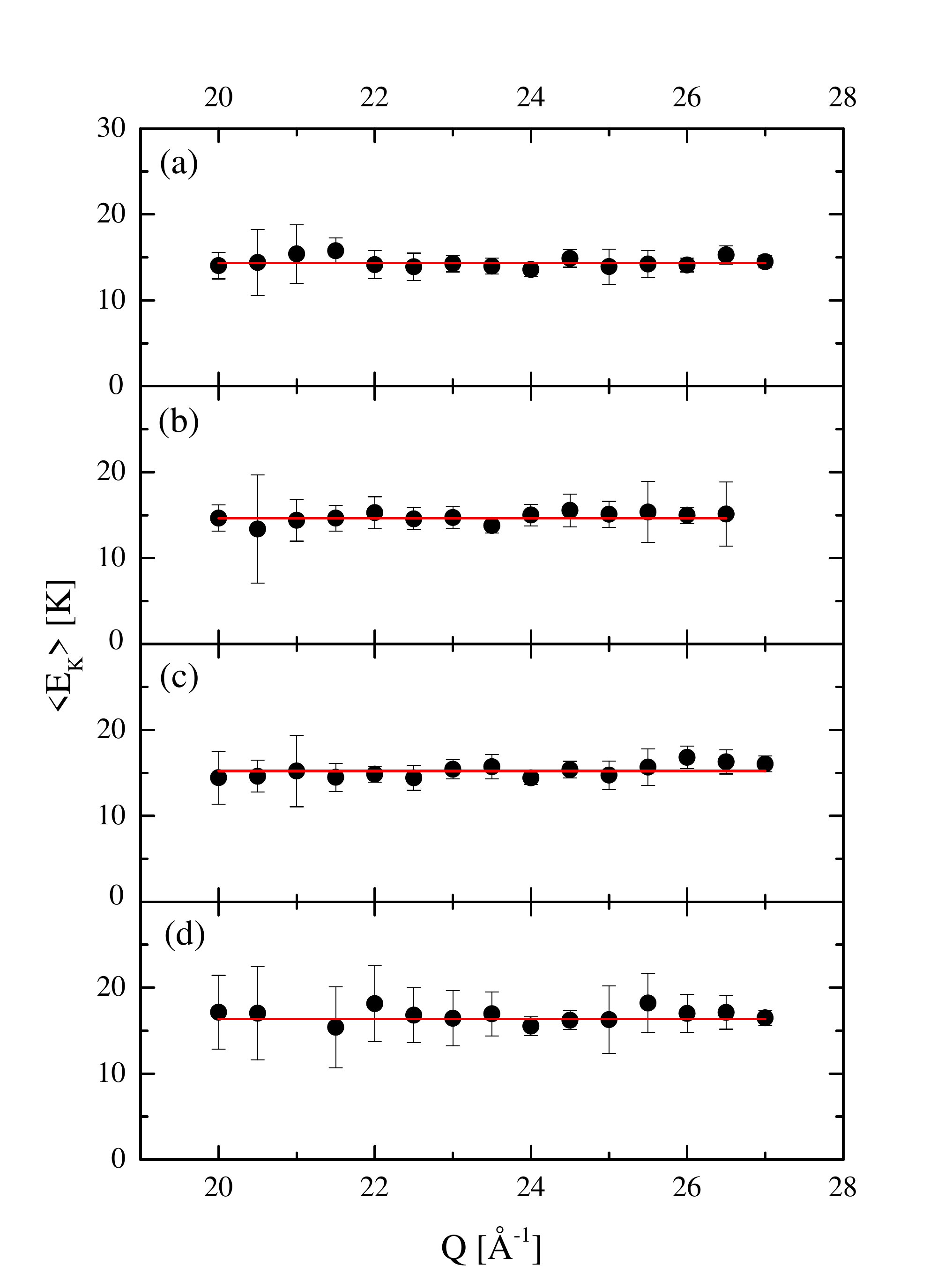}
\caption{Experimental estimates for the average atomic kinetic energy $\langle E_K\rangle$ obtained from the phenemonological model $J_{IA}(Y)$: (a) 1.090(10) K, (b) 1.800(4) K, (c) 2.100(4), and (d) 2.650(15) K.  The best estimate for $\langle E_K\rangle$ at each temperature is shown by a horizontal red line.
\label{fig:KEq}}
\end{figure}

Figure \ref{fig:KEq} illustrates the kinetic energies $\langle E_K\rangle$ extracted from $J(Y, Q)$ as a function of $Q$.  The observed kinetic energy is constant with $Q$, as required by the $\omega^2$-sum rule.  The best estimate for the kinetic energy $\langle E_K\rangle$ is obtained by combining the results of these measurements at each $Q$ by means of a weighted average.  Experimental estimates for the average kinetic energy $\langle E_K\rangle$ are listed in Table \ref{table:KE}.  Equivalent results for $\langle E_K\rangle$ are obtained when $J_{\textrm{IA}}^{(P)}(Y)$ is broadened only by the instrumental resolution function and when it is broadened by both the resolution and final state effects.  This is due to the fact that $\langle E_K\rangle$ is determined by only the intrinsic (i.e. resolution corrected) second moment of the scattering.  Therefore, these empirical estimates may be viewed as model-independent.

\begin{figure}
\includegraphics[width=\textwidth, height=\textheight, keepaspectratio]{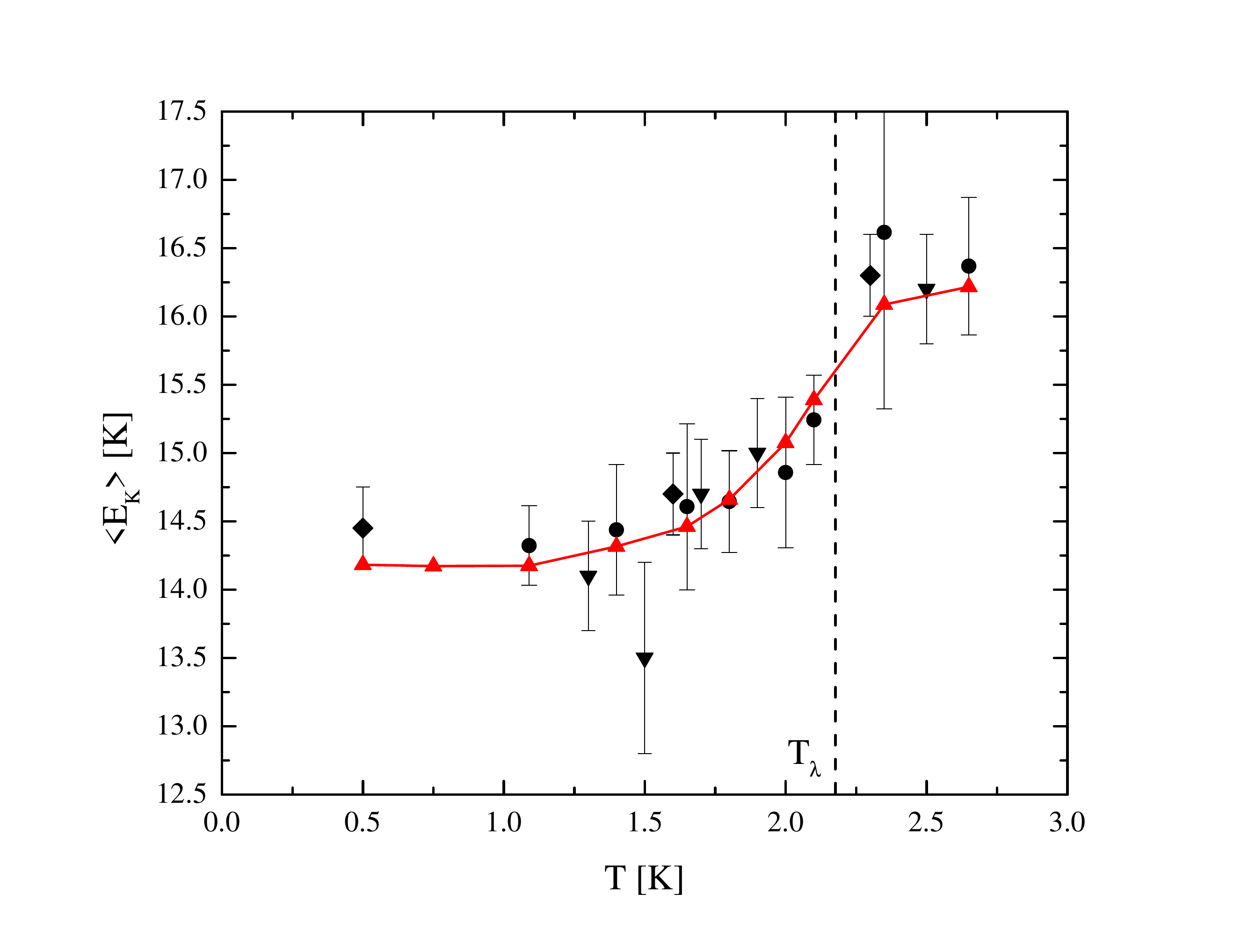}
\caption{The average atomic kinetic energy $\langle E_K\rangle$ of liquid $^4$He under saturated vapor pressure.  Experimental estimates: present ARCS study (circles), MARI\cite{Azuah, GAS} (diamonds), and eVS\cite{Mayers} (triangles).  Our QMC predictions are shown as red triangles, the line being a guide to the eye.
\label{fig:KE}}
\end{figure}

Theoretical and experimental values for the average kinetic energy $\langle E_K\rangle$ of liquid $^4$He under SVP are shown in Figure \ref{fig:KE}.  The QMC calculations predict that $\langle E_K\rangle$ increases from $14.17(2) \textrm{ K}$ at 1.09 K to $15.39(5) \textrm{ K}$  at 2.100 K.  The kinetic energy increases rapidly through the superfluid phase transition at $T_\lambda$, reaching a relatively constant value of $\approx16.2$ K in the normal liquid.  The QMC calculations are in excellent agreement with the ARCS data set presented in this paper, as well as previous investigations using the MARI\cite{Azuah, GAS} and eVS\cite{Mayers} spectrometers.

The measured scattering from liquid $^4$He is consistent with many possible forms for the momentum distribution $n(k)$.  In Section \ref{sub:lineshape}, we showed that \emph{ab initio} calculations of $J(Y, Q)$ are in agreement with the observed scattering.  These calculations predict a finite Bose condensate fraction $n_0$ in the superfluid phase.  However, the observed scattering is also consistent with models that do not incorporate a Bose condensate in the superfluid phase, such as the phenemonological model.  The problem of inverting the scattering data $J(Y, Q)$ to a unique momentum distribution $n(k)$ is ill-posed\cite{Sivia}.  No information about the Bose condensate fraction $n_0$ can be obtained from the neutron Compton scattering data without the help of theoretical models.

\subsection{Empirical Estimates of $n_0$}
In this section, we obtain empirical estimates of the Bose condensate fraction $n_0$ as a function of temperature.  We introduce two different parameterized expressions for the IA-scattering $J_{\textrm{IA}}(Y)$ that explicitly incorporate a Bose condensate.  Both models are broadened by instrumental resolution $I(Y, Q)$ and final state effects $R_{\textrm{CK}}(Y, Q)$ when fitting the scattering data.

\textbf{Model A: Expansion in Orthogonal Polynomials.} The first model represents the momentum distribution $n(k)$ as the sum of a $\delta$-function singularity plus a non-Gaussian peak\cite{Andreani}.
\begin{equation}
n(\mathbf{k}) = n_0\delta(\mathbf{k}) + (1-n_0)\frac{e^{-k^2/2\sigma^2}}{(2\pi\sigma^2)^{3/2}}\left(1 + \sum_{n = 2}^\infty a_n (-1)^n L_n^{1/2}\left(\frac{k^2}{2\sigma^2}\right) \right)
\end{equation}
Here $L_n^{1/2}$ is an associated Laguerre polynomial of order $n$.  
When transformed into the $Y$-coordinates, Model A has the following form:
\begin{equation}\label{Hermite}
J_{IA}(Y) = n_0\delta(Y - Y_c) + (1 - n_0)\frac{e^{-(Y-Yc)^2/2\sigma^2}}{\sqrt{2\pi\sigma^2}}\left[1 + \sum_{n = 2}^{\infty} a_n \frac{1}{2^{2n}n!}H_{2n}\left(\frac{Y - Y_c}{\sigma\sqrt{2}}\right)  \right] + aY + b
\end{equation}
Here $H_n$ is the Hermite polynomial of order $n$.  An overall scale factor is also included to allow for uncertainty in the absolute intensity scale.  The second moment of the scattering is equal to $(1 - n_0)\sigma^2$.  Nonzero values of the expansion coefficients $\{a_n\}$ do not affect the second moment of $J_{\textrm{IA}}(Y)$.  Again, we include a linear background to account for residual multiple scattering in the tails of $J(Y, Q)$.

When fitting the scattering data, we have kept $n_0$, $\sigma$, and as few expansion coefficients $\{a_n\}$ needed to obtain a $\chi^2$ of approximately unity.  Only terms up to $a_4$ were kept.

\textbf{Model B: Cumulant Expansion.}  The second model represents the momentum distribution $n(k)$ in terms of a cumulant expansion\cite{Glyde}.  The momentum distribution is expressed as a sum of three terms:
\begin{equation}
n(\mathbf{k}) = n_0(\delta(\mathbf{k}) + f(k)) + A_1 n^*(k)
\end{equation}
The first term is the $\delta$-function singularity of the condensate itself.  The second term, $n_0f(k)$, is the weaker singularity produced by the coupling of virtual phonons with the condensate.  
\begin{equation}
f(k) = \frac{1}{(2\pi)^3}\frac{mc}{2\hbar \rho}\frac{1}{k}\left[2N(c\hbar k) + 1\right]e^{-k^2/k_c^2}.
\end{equation}
Here $m$ is the mass of a helium atom; $\rho$ is the number density of the liquid; $c$ is the phonon velocity; and $N$ is the Bose population factor.  If $\hbar ck\gg k_BT$, then $f(k)$ is proportional to $1/k$; if $\hbar ck\ll k_BT$, then $f(k)$ is proportional to $1/k^2$.  The exponential $e^{-k^2/k_c^2}$ is introduced \emph{ad hoc} to smoothly cut off the contribution of $f(k)$ outside the phonon region.  Following the literature\cite{Glyde}, we fix $k_c = 0.5 \textrm{ \AA}^{-1}$.

The third term $A_1 n^*(k)$ is the momentum distribution of the atoms above the condensate.  They are described by a cumulant expansion:
\begin{equation}
\widetilde{n}^*(s) = \exp\left[\sideset{}{'}\sum_{n = 2}^{\infty} \alpha_{n}\frac{(is)^{n}}{n!} \right] \approx \exp\left[-\frac{\alpha_{2}s^{2}}{2!} +\frac{\alpha_{4}s^{4}}{4!}-\frac{\alpha_{6}s^{6}}{6!}\right]
\end{equation}
Here $\widetilde{n}^*(s)$ is the Fourier transform of $n^*(k)$.  The prime indicates that only terms with even $n$ contribute.  The coefficients ${\alpha_n}$ of the expansion are the statistical cumulants of $n^*(k)$.  

There is no simple analytic expression for $J_{IA}(Y)$ for this model of $n(k)$ when terms up to $\alpha_6$ are retained.
\begin{equation}\label{eq:ModelB}
J_{\textrm{IA}}(Y) = n_0(\delta(Y - Y_c) + \overline{f}(Y - Y_c)) + A_1\overline{J}(Y- Y_c) + aY + b
\end{equation}
Here the overbar signifies the result of transforming $f(k)$ and $n^*(k)$ into the $Y$-scaling variable.  The adjustable parameters describing the momentum distribution are: $n_0$, $\alpha_2$, $\alpha_4$, and $\alpha_6$.

\textbf{Overlap between the models.} These two models for $J_{\textrm{IA}}(Y)$ appear to treat the uncondensed part of the momentum distribution $n(k)$ very differently.  However, there is a special case where they are exactly equivalent.  If the higher order cumulants are small, then then $\widetilde{n}^*(s)$ can be approximated as follows:
\begin{equation}
\widetilde{n}^*(s) \approx \exp\left[-\frac{\alpha_{2}s^{2}}{2}\right]\left(1 +\frac{\alpha_{4}s^{4}}{4!}\right)
\end{equation}
In this particular case, the characteristic function $\widetilde{n}^*(s)$ transforms analytically:
\begin{equation}\label{delta}
J_{IA}^*(Y) = \frac{1}{\sqrt{2\pi\sigma^2}}\exp\left[-\frac{Y^2}{2\sigma^2}\right]\left(1 + \frac{\delta}{8}\left(1 - \frac{2Y^2}{\sigma^2} + \frac{Y^4}{3\sigma^4}\right)\right)
\end{equation}
Here $\sigma^2 = \alpha_2$ and $\delta = \alpha_4/\alpha_2^2$.  This expression, Equation \ref{delta}, is equivalent to keeping only $H_4$ in Equation \ref{Hermite} with $\delta = 3a_2$.

We have used the full expression in Equation \ref{eq:ModelB} to fit the scattering in almost all of the data sets.  However, given the statistical noise in the 1.40 K data measurements, we have used Equation \ref{delta} to represent the uncondensed part of $n(k)$.

\begin{figure}
\includegraphics[width=\textwidth, height=\textheight, keepaspectratio]{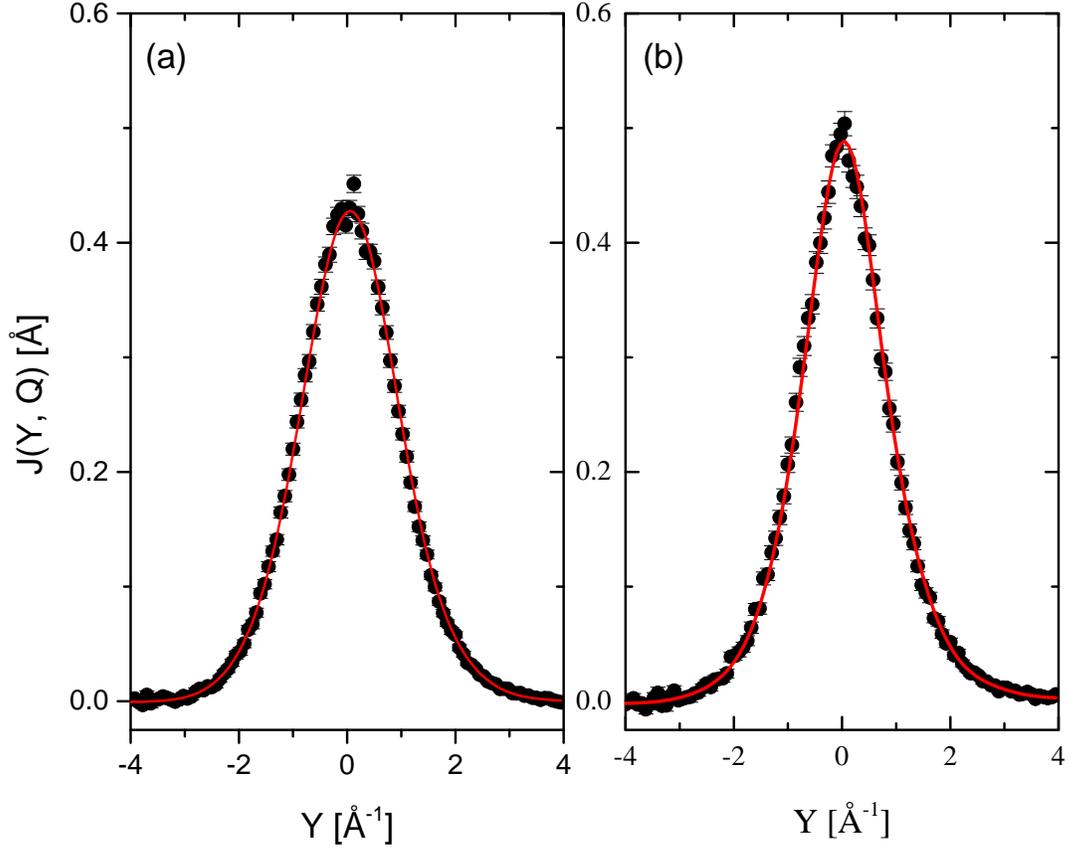}
\caption{The observed scattering for $Q = 26.5 \textrm{ \AA}^{-1}$ is fit to Model B at (a) 2.650(10) K and (b) 1.090(10) K.  From the fit shown in panel (a), we find that $n_0 = 0.000(9)$, $\alpha_2 = 0.90(2) \textrm{ \AA}^{-2}$, and $\chi^2$ is 0.975 at this $Q$.  Meanwhile, the fit shown in panel (b) yields $n_0 = 0.068(7)$, $\alpha_2 = 0.87(3) \textrm{ \AA}^{-2}$, and $\chi^2$ is 0.866 for this $Q$.
\label{fig:CEfit}}
\end{figure}

\textbf{Results of the Fits.} We fit the scattering data $J(Y, Q)$ at all temperatures $T$ using Model A and Model B, with the exception of the 2.35 K data set, where the statistical precision of the data is too low to obtain a meaningful estimate of $n_0$.   Figure \ref{fig:CEfit} shows representative fits to the scattering data using Model B.  Typical values of $\chi^2$ are close to unity and the residuals do not indicate any systematic discrepancies between the model curve and the scattering data.

\begin{figure}
\includegraphics[width=\textwidth, height=\textheight, keepaspectratio]{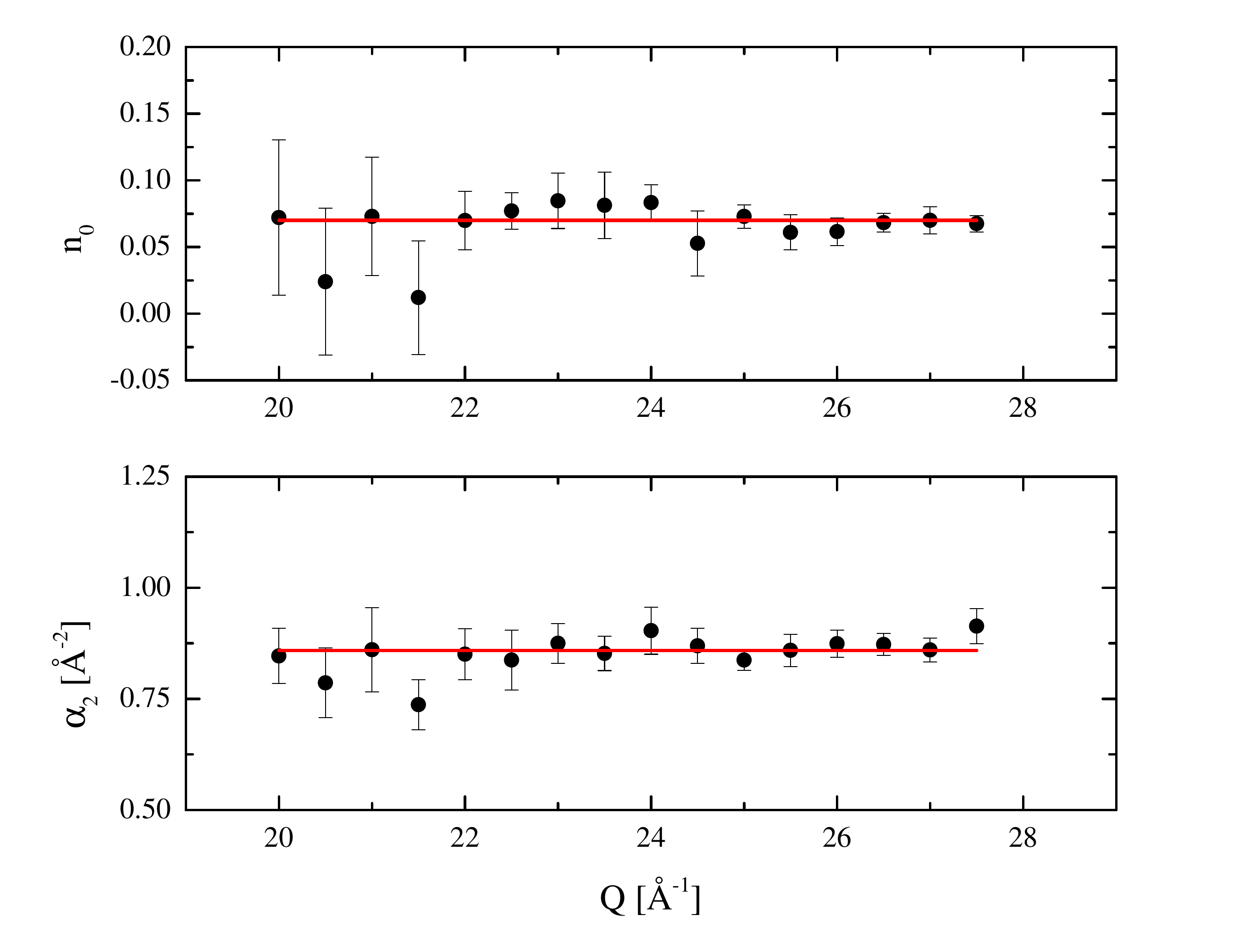}
\caption{Experimental estimates for the condensate fraction $n_0$ and and the second cumulant $\alpha_2$ obtained from the $T = 1.090(10) \textrm{ K}$ data set using Model B described in the text.  The best estimate for each quantity is shown by a horizontal red line.
\label{fig:CEq}}
\end{figure}

Figure \ref{fig:CEq} plots the values of $n_0$ and $\alpha_2$ obtained at $T = 1.090(10) \textrm{ K}$.  The observed values of the condensate fraction $n_0$ and second cumulant $\alpha_2$ are independent of $Q$.  The best estimate for these quantities is obtained by taking a weighted average over all values of $Q$.  For this temperature, we obtained $n_0 = 0.070(4)$ and $
\alpha_2 = 0.859(10) \textrm{ \AA}^{-2}$.  The average kinetic energy $\langle E_K\rangle$ is $14.2(2)$, which is consistent with the result of the phenomenological model.  

Empirical estimates for the condensate fraction $n_0$ obtained from the ARCS data set are listed in Table \ref{table:n0_ARCS}.  We have also carried out the same analysis upon the PHOENIX data sets\cite{SosnickLong, SnowLong}.  These results are shown in Table \ref{table:n0_PHOENIX}.  Consistent results are obtained from the ARCS and PHOENIX data sets when common models are used to analyze the scattering data.

\begin{figure}
\includegraphics[width=\textwidth, height=\textheight, keepaspectratio]{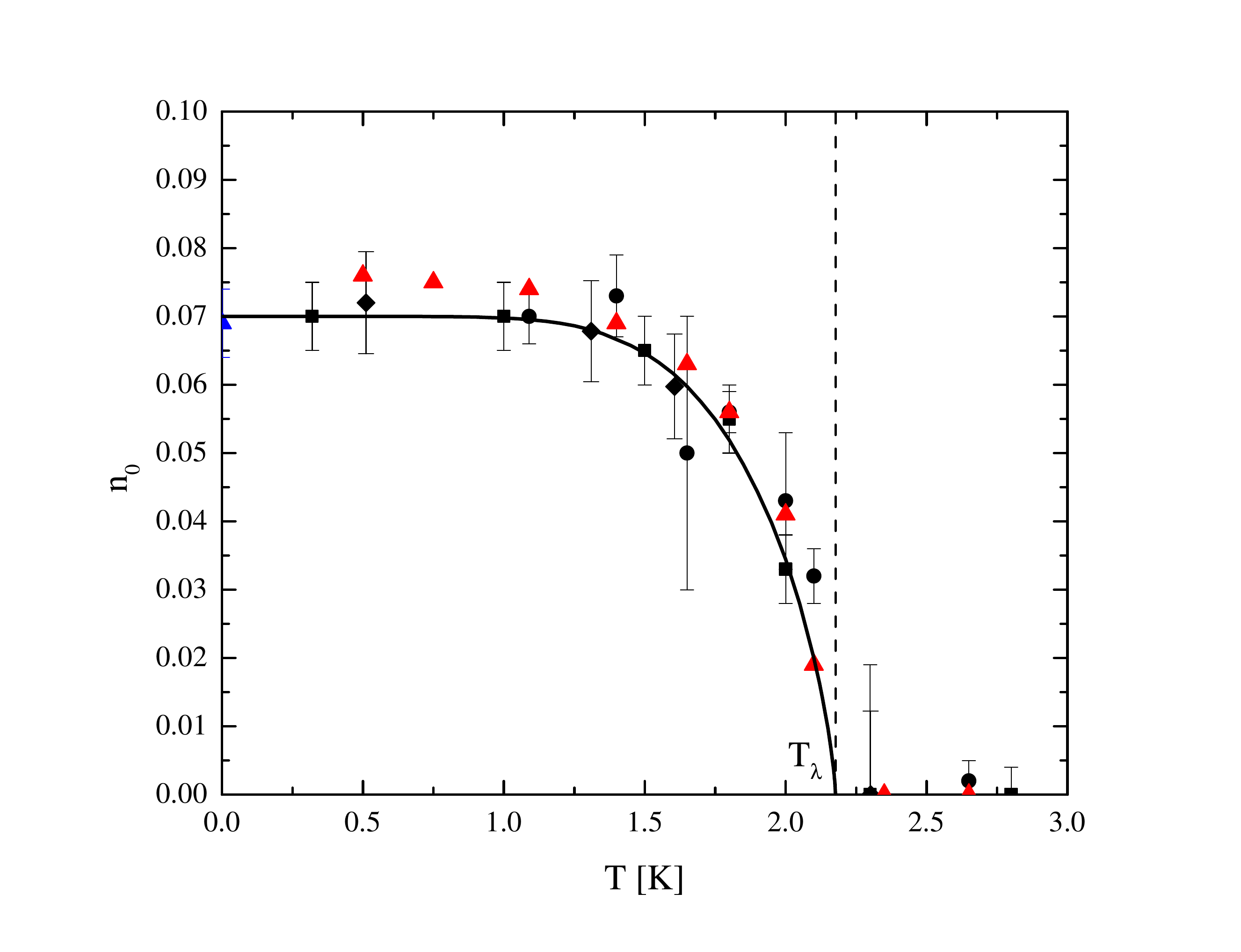}
\caption{The Bose condensate fraction $n_0$ of liquid $^4$He under saturated vapor pressure.  Experimental estimates: present ARCS study (circles), MARI (diamonds), and our re-analysis of the PHOENIX data set (squares).  Theoretical points: current QMC estimates (red triangle), Reptation Quantum Monte Carlo (blue triangle), and the GPS theory (solid black line).
\label{fig:n0}}
\end{figure}

Figure \ref{fig:n0} plots empirical estimates of $n_0$ at saturated vapor pressure obtained from the ARCS, MARI, and PHOENIX spectrometers.  These values are compared with our present (QMC) estimates for $n_0$, which are obtained by averaging the value of the one-body density matrix $\widetilde{n}(s)$ at distances above $7 \textrm{ \AA}$.  Also shown is a ground state QMC prediction\cite{moroni}.  At low temperatures, the condensate fraction is close to $7.5\%$.  No significant temperature dependence is observed below 1.1 K.  However, above 1.1 K, the condensate fraction $n_0$ decreases rapidly toward zero as the transition temperature $T_\lambda$ is approached.  In the normal fluid phase, the condensate fraction $n_0$ is zero.

The relationship between the phonon-roton spectrum and the Bose condensate in superfluid $^4$He is presently an open question.  Giorgini, Pitaevskii, and Stringari\cite{GPS} proposed that the thermal excitation of rotons is the chief mechanism driving the depletion of the condensate as the temperature approaches $T_\lambda$.  According to their theory, the ground state value of the condensate fraction $n_0(0)$ is driven by the smallness of the ratio $3k_BT_\lambda m/\hbar^2Q_R^2 \approx 0.15$, where $Q_R$ is the roton wavevector.  If $\rho_n(T)$ is the normal fluid fraction at temperature $T$, then they predict that the temperature dependence of the condensate fraction $n_0(T)$ is:
\begin{equation}\label{eq:GPS}
n_0(T) = n_0(0) \left(1 - \frac{T}{T_\lambda}\rho_n(T)\right)
\end{equation}
The solid line in Figure \ref{fig:n0} is obtained by setting $n_0(0)$ equal to $7\%$.  There is good agreement between the experimental data and the predictions of Equation \ref{eq:GPS}.

The estimated values of $n_0$ obtained from the eVS (now VESUVIO) instrument, a nuclear resonance foil spectrometer, have not been shown in Figure \ref{fig:n0}.  The eVS group\cite{Mayers} reports that $n_0$ is zero at 2.5 K.  They also claim that the condensate fraction $n_0$ increases from $0.010(4)$ at 1.9 K to $0.015(4)$ at 1.3 K.  Their values for $n_0$ in the superfluid phase are inconsistent with the ARCS, PHOENIX, and MARI data sets, as well QMC predictions, at the level of $1\sigma$.  We believe that the origin of this discrepancy is the comparatively coarse energy resolution that was available to the eVS group.  For example, those authors note that the use of the U-foil analyzer produces a resolution function $I(Y, Q)$ having a central Gaussian width of $1.53 \textrm{ \AA}^{-1}$ and Lorentzian tails of width $1.46 \textrm{ \AA}^{-1}$ when $Q = 152 \textrm{ \AA}^{-1}$.  This excludes the possibility of a detailed lineshape analysis, as the intrinsic width of the scattering is approximately $2 \textrm{ \AA}^{-1}$.

\section{Conclusions}

In this paper, we presented a new high-resolution neutron Compton scattering study of liquid $^4$He under saturated vapor pressure.  The measurements were performed using the ARCS spectrometer at the Spallation Neutron Source.  We found that there is excellent agreement between the observed neutron Compton profile $J(Y, Q)$ and \emph{ab initio} predictions of its lineshape.  Model fit functions were used to obtain empirical estimates for the average atomic kinetic energy $\langle E_K\rangle$ and Bose condensate fraction $n_0$ as a function of temperature.  These quantities are also in excellent agreement with \emph{ab initio} calculations.  Finally, by a reanalysis of the PHOENIX data, we have resolved an apparent contradiction in the literature over the magnitude of the condensate fraction $n_0$.  We conclude that the scattering data provides compelling evidence for the existence of a Bose condensate in superfluid $^4$He.

\begin{acknowledgments}
The authors are grateful to David Sprinkle of Indiana University for lending his expertise to the design, construction, and testing of the cyrogenics built for this study.  Saad Elorfi and Mark Loguillo of the Spallation Neutron Source provided technical support for this experiment.  We also recognize helpful scientific discussions with Doug Abernathy, Richard Azuah, Jiao Lin, Matthew Stone, and Peter Willendrup.  This report was prepared, in part, by Indiana University under award 70NANB10H255 from the National Institute of Standards and Technology.  Matthew S. Bryan acknowledges support under NSF grant DGE-1069091.  This research at ORNL's Spallation Neutron Source was sponsored by the Scientific User Facilities Division, Office of Basic Energy Sciences, U.S. Department of Energy, as well as by the Natural Science and Engineering Research Council of Canada.  Computing support from Westgrid is gratefully acknowledged.
\end{acknowledgments}
\newpage

 \begin{table}%[H] add [H] placement to break table across pages
\caption{Values of the average atomic kinetic energy $\langle E_K \rangle$ estimated from the ARCS data set.
\label{table:KE}}
 \begin{ruledtabular}
 \begin{tabular}{|c c c|}
$T$ [K] & $\langle E_K\rangle$ [K] & QMC [K]\\ %[0.5ex]
\hline 
1.090(10) & 14.3(3) & 14.17(2)\\ 
1.400(10) & 14.4(5) & 14.32(2) \\
1.650(4) & 14.6(6) & 14.46(2) \\
1.800(4) & 14.6(4) & 14.66(2) \\
2.000(4) & 14.9(6) & 15.08(3) \\ 
2.100(4) & 15.2(3) & 15.39(5) \\
2.35(3) & 16.6(1.3) & 16.09(2) \\
2.650(15) & 16.4(5) & 16.22(1) \\
 \end{tabular}
 \end{ruledtabular}
 \end{table}

 \begin{table}%[H] add [H] placement to break table across pages
\caption{Values of the condensate fraction $n_0$ estimated from the ARCS data set.
\label{table:n0_ARCS}}
 \begin{ruledtabular}
 \begin{tabular}{|c c c c|}
$T$ [K] & Model A & Model B & QMC\\ %[0.5ex]
\hline 
1.090(10) & 0.073(2) & 0.070(4) & 0.075(2) \\ 
1.400(10) & 0.071(6) & 0.073(6) & 0.069(2) \\
1.650(4) & 0.051(13) & 0.05(2) & 0.063(2) \\
1.800(4) & 0.061(3) & 0.056(3) & 0.056(2)\\
2.000(4) & 0.039(6) & 0.043(10) & 0.041(2)\\ 
2.100(4) & 0.034(3) & 0.032(4) & 0.019(2)\\
2.35(3) & --      & -- & 0 \\ 
2.650(15) & 0.000(1) & 0.002(3) & 0\\
 \end{tabular}
 \end{ruledtabular}
 \end{table}

 \begin{table}%[H] add [H] placement to break table across pages
\caption{Values of the condensate fraction $n_0$ estimated from the PHOENIX data set.
\label{table:n0_PHOENIX}}
 \begin{ruledtabular}
 \begin{tabular}{|c c c|}
$T$ [K] & Model A & Model B\\ %[0.5ex]
\hline 
0.32 & 0.071(9) & 0.070(5) \\ 
1.00 & 0.069(9) & 0.070(5) \\
1.50 & 0.069(9) & 0.065(5) \\
1.80 & 0.045(9) & 0.055(5) \\ 
2.00 & 0.042(14) & 0.033(5) \\
2.30 & 0.000(2)  & 0.000(19) \\
2.80 & 0.001(4) & 0.000(4) \\
3.50 & 0.000(1) & 0.000(11) \\
 \end{tabular}
 \end{ruledtabular}
 \end{table}

\newpage

\clearpage

\bibliography{DINSpaper_v4.bib}

\begin{thebibliography}{77}
\bibitem [{azi()}]{aziznote}%
  \BibitemOpen
  \href@noop {} {}\bibinfo {note} {Revisions of the original Aziz potential
  have been published over the past few decades, but the differences are
  relatively small, and do not affect the vaue of the condensate fraction to a
  significant degree. See, for instance, Ref. \onlinecite{moroni}.}\BibitemShut
  {Stop}%
\end{thebibliography}

\end{document}